\documentclass[aps,prb,superscriptaddress,reprint,floatfix]{revtex4-2}
\usepackage{graphicx}
\usepackage{dcolumn}
\usepackage{bm}
\usepackage{amsmath}
\usepackage{float} 
\usepackage[export]{adjustbox}
\usepackage{xcolor}
\usepackage[frozencache=true,cachedir=.]{minted}
\usepackage[utf8]{inputenc}
\usepackage{textgreek}
\definecolor{light-gray}{gray}{0.95}
\usepackage{bm,graphicx,hyperref}
\hypersetup{%
  breaklinks = {true},
  citecolor = {blue},
  colorlinks = {true},
  linkcolor = {red},
}

\begin{document}
\title{Numerical package for QFT calculations of defect-induced phenomena in graphene}

\author{Hillol Biswas}
\thanks{These two authors contributed equally}
\affiliation{Department of Physics, National University of Singapore, 2 Science Drive 3, 117542, Singapore}
\affiliation{Centre for Advanced 2D Materials, National University of Singapore, 6 Science Drive 2, 117546, Singapore}

\author{Harshitra Mahalingam}
\thanks{These two authors contributed equally}
\affiliation{Institute for Functional Intelligent Materials, National University of Singapore, 4 Science Drive 2, 117544, Singapore}

\author{Aleksandr Rodin}
\affiliation{Centre for Advanced 2D Materials, National University of Singapore, 6 Science Drive 2, 117546, Singapore}
\affiliation{Yale-NUS College, 16 College Avenue West, 138527, Singapore}

\date{\today}

\begin{abstract}

We introduce a computationally efficient method based on the path integral formalism to describe defect-modified graphene. By taking into account the entire Brillouin zone, our approach respects the lattice symmetry and can be used to investigate both short-range and long-range effects. The proposed method's key advantage is that the computational complexity does not increase with the system size, scaling, instead, with the number of defects. As a demonstration of our method, we explore the graphene-mediated RKKY interaction between multiple magnetic impurities. Our results concur with earlier findings by showing that the interaction strength and sign depend on various factors like impurity separation, sublattice arrangement, and system doping. We demonstrate that frustration can be introduced between the impurity spins by controlling their relative positions and that this frustration can be switched on and off by tuning the chemical potential of the system.

\end{abstract}

\maketitle

\section{Introduction}

Theoretical studies of defects in graphene typically take one of three approaches: DFT simulations for atomically-precise short-range features,~\citep{Yazyev2007, Boukhvalov2008, Gerber2010afs, Park2012msa, Park2013spe, Frank2017} exact diagonalization of the tight-binding Hamiltonian for large but finite systems, where retaining the lattice structure is essential,~\citep{Garcia-Lastra2010sdb, Black-Schaffer2010rci, Jung2013} and the computationally-efficient $\mathbf{k}\cdot\mathbf{p}$ Dirac Hamiltonian to describe long-range and low-energy phenomena.~\citep{Pereira2007, Shytov2007, Shytov2007a, Uchoa2008, CastroNeto2009, Shytov2009lri,Kogan2011rii, Lebohec2014art, Agarwal2017lre,Frank2017, Agarwal2019pas} Over the past decade, advances in graphene fabrication and manipulation have made the highly-controlled experimental investigation of atomic-scale phenomena possible. For example, atomically precise deposition of adsorbates has allowed researchers to explore the role of impurity interaction in magnetism~\citep{GonzalezHerrero2016asc}, study electronic scattering due to individual impurities~\citep{Brar2011}, and confirm the supercritical potential regime predicted theoretically~\citep{Wang2013, Lu2019}. Moreover, individual-atom doping~\citep{Telychko2014, Telychko2015} made it possible to observe the Berry phase in the presence of single nitrogen atoms~\citep{Dutreix2019}, induce a controlled migration of silicon dopants~\citep{Tripathi2018ebm}, and shed light on the effects of single dopants on the electronic structure of the host material.~\citep{Telychko2021a}

In a significant portion of experimental studies, highly localized perturbations give rise to spatially-extended features. Consequently, one may deem the exact diagonalization an ideal approach as DFT becomes computationally infeasible because of the large supercell requirements and the Dirac Hamiltonian fails to capture the appropriate structure close to the perturbation. There is, however, a caveat: the system used in the exact diagonalization calculations must be sufficiently large to avoid finite-size effects. Including multiple spatially separated defects increases the minimum system size as one needs to make sure that the system edges are far enough away from all the perturbations.

An approach that does not lead to a drastic increase in computational complexity with additional defects while also respecting the lattice symmetry involves the field-theoretic formulation of the problem using the full tight-binding Hamiltonian instead of the simplified $\mathbf{k}\cdot\mathbf{p}$ version. This method has been used to, for example, study the effects of individual hydrogen adsorbates~\citep{Noori2020, Noori2020a}. Unfortunately, despite the utility of this approach, it remains isolated from the experimental community for which it would be the most useful, partly because of the perceived difficulty of QFT. Moreover, even for the community members familiar with the formalism, the time and effort required to set up the computational pipeline using this approach are substantial because code from previous studies is either not readily available or not fit for use by outside parties. As a result, there is a lot of redundant effort in the community, slowing down the research progress.

In this work, we introduce GrapheneQFT.jl,~\citep{rodinalex_harshitra-m_2022} an extendable package written in JULIA programming language~\citep{Bezanson2017} designed to calculate a variety of experimentally relevant QFT quantities in graphene in the presence of external perturbations. In particular, this package can compute electronic density, Green's and spectral functions, and free energy in a graphene system containing adsorbates, dopants, and local gating. The paper is organized in the following manner. In Sec.~\ref{sec:Model}, we introduce the model and derive the relevant expressions to be used in the calculations. This is an analytical section and the reader more interested in the applications of the package can move directly to Sec.~\ref{sec:Results}, where we demonstrate the use of the package. Specifically, we focus here on the interaction between spin impurities in graphene via the Ruderman-Kittel-Kasuya-Yoshida
(RKKY) coupling.~\cite{PhysRev.96.99, 10.1143/PTP.16.45, PhysRev.106.893} We investigate the sign dependence of the RKKY interaction on the system parameters and the impurity arrangement.~ \citep{Saremi2007rih, Black-Schaffer2010rci,Agarwal2017lre} Summary and outlook are provided in Sec.~\ref{sec:Summary}.
\section{Model}
\label{sec:Model}

\subsection{Two-Component System}
\label{sec:Two_Component_System}

Instead of starting directly with the problem of defects in graphene, we begin by focusing on a more general scenario which will make the derivation more transparent. Consider two quantum systems described by single-particle second-quantized Hamiltonians $\sum_{jk} c^\dagger_j H_{jk}c_k$ and $\sum_{jk} d^\dagger_j h_{jk}d_k$, where $j$ and $k$ label the single-particle states in the systems. We will refer to the first system as ``bulk" to indicate that it contains a large number of states compared to the second one, which we will label the ``impurity" system.

Next, we introduce a coupling between the two systems and a perturbation that modifies the matrix elements for the bulk, leading to

\begin{align}
    \hat{\mathcal{H}} &= \sum_{jk} c^\dagger_j H_{jk}c_k 
    + 
    \sum_{jk} g^\dagger_j h_{jk}g_k
    \nonumber
    \\
    &+
    \sum_{jk} \left(c^\dagger_j V_{jk}g_k + g^\dagger_k V_{jk}^* c_j\right)
    +
    \sum_{jk} c^\dagger_j \Delta_{jk}c_k
    \nonumber
    \\
    &= \mathbf{c}^\dagger H\mathbf{c} + \mathbf{c}^\dagger \Delta \mathbf{c}
    +\left(\mathbf{c}^\dagger V\mathbf{g} + \mathbf{g}^\dagger V^\dagger \mathbf{c}\right) + \mathbf{g}^\dagger h\mathbf{g}\,.
    \label{eqn:H}
\end{align}
The coupling is given by the first term of the second line, while the bulk perturbation is the second term in the second line. As the final step, we made the Hamiltonian more compact by writing the sums as products of coupling matrices and vectors of operators.

The normal-ordered Hamiltonian in Eq.~\eqref{eqn:H} can be transcribed into the imaginary-time action
\begin{align}
    S &= \sum_{n}\Big[ \bar{\boldsymbol{\psi}}_n \overbrace{\left(-i\omega_n - \mu + H\right)}^{-G^{-1}_{i\omega_n + \mu}}\boldsymbol{\psi}_n + \bar{\boldsymbol{\psi}}_n\Delta \boldsymbol{\psi}_n
    \nonumber
    \\
    &+\left(\bar{\boldsymbol{\psi}}_n V\boldsymbol{\phi}_n + \bar{\boldsymbol{\phi}}_n V^\dagger \boldsymbol{\psi}\right) + \bar{\boldsymbol{\phi}}_n \underbrace{\left(-i\omega_n - \mu + h\right)}_{-\Gamma_{i\omega_n+\mu}^{-1}}\boldsymbol{\phi}_n\Big]
    \nonumber
    \\
    &=
    \sum_n
    \begin{pmatrix}
        \bar{\boldsymbol{\psi}}_n & \bar{\boldsymbol{\phi}}_n
    \end{pmatrix}
    \begin{pmatrix}
        -G_{i\omega_n+\mu}^{-1} + \Delta & V
        \\
        V^\dagger & -\Gamma_{i\omega_n+\mu}^{-1}
    \end{pmatrix}
    \begin{pmatrix}
        \boldsymbol{\psi}_n \\ \boldsymbol{\phi}_n
    \end{pmatrix}\,,
    \label{eqn:S}
\end{align}
where $\omega_n$ are the fermionic Matsubara frequencies, $\mu$ is the chemical potential, and $\boldsymbol{\phi}_n$ and $\boldsymbol{\psi}_n$ ($\bar{\boldsymbol{\phi}}_n$ and $\bar{\boldsymbol{\psi}}_n$) are vectors of Grassmann numbers corresponding to $\mathbf{g}$ and $\mathbf{c}$ ($\mathbf{g}^\dagger$ and $\mathbf{c}^\dagger$). We identify $G_{z}$ and $\Gamma_{z}$ as the Green's functions for the two isolated and unperturbed systems. The matrix form of the action makes it straightforward to calculate the partition function by exponentiating $-S$ and integrating over all the Grassmann variables:

\begin{align}
    \mathcal{Z} &= \prod_n
    \Bigg|\beta
    \underbrace{\begin{pmatrix}
        -G_{i\omega_n+\mu}^{-1} + \Delta & V
        \\
        V^\dagger & -\Gamma_{i\omega_n+\mu}^{-1}
    \end{pmatrix}}_{-\mathbf{G}_{i\omega_n+\mu}^{-1}}
    \Bigg|\,,
    \label{eqn:Z}
\end{align}
where $\beta = 1 / (k_BT)$. The quantity $\mathbf{G}_{z}$ is the full Green's function for the composite system, given explicitly by

\begin{equation}
    \mathbf{G}_{z} 
    =
    \begin{pmatrix}
         \mathcal{G}_z &\ \mathcal{G}_z V \Gamma_{z}
        \\
         \Gamma_{z}V^\dagger \mathcal{G}_z & \Gamma_{z}+ \Gamma_{z}V^\dagger \mathcal{G}_z V \Gamma_{z}
    \end{pmatrix}\,,
    \label{eqn:G_full}
\end{equation}
where
\begin{align}
    \mathcal{G}_z 
    &=\left(G_{z}^{-1} - \Delta - V\Gamma_{z} V^\dagger\right)^{-1} 
    \nonumber
    \\
    &= G_{z}+G_{z}\left(\Delta + V\Gamma_{z} V^\dagger\right)
    \nonumber
    \\
    &\times \left[1-G_{z}\left(\Delta + V\Gamma_{z} V^\dagger\right)\right]^{-1}G_{z} \,,
    \label{eqn:G_full_bulk}
\end{align}
is the full Green's function of the bulk and the bottom right block in Eq.~\eqref{eqn:G_full} corresponds to the full Green's function of the impurity states including their coupling to the perturbed bulk system.

From Eq.~\eqref{eqn:G_full_bulk}, one sees that $\mathcal{G}_z$ is comprised of two parts: the pristine bulk system $G_z$ and the perturbation-induced correction term, which we denote $\delta \mathcal{G}_z$. The matrix elements of $\delta \mathcal{G}_z$ are 
\begin{align}
    \delta \mathcal{G}^{jk}_z &= \sum_{lm} G^{jl}_{z} \Bigg\{ \left(\Delta + V\Gamma_{z} V^\dagger\right) 
    \nonumber
    \\
    &\times \left[1-G_{z}\left(\Delta + V\Gamma_{z} V^\dagger\right)\right]^{-1} \Bigg\}_{lm} G^{mk}_{z} \,,
    \label{eqn:deltaG_term}
\end{align}
where the sum over $l$ and $m$ includes all the states in the bulk system. This expression can be made considerably simpler by rewriting $\Delta + V\Gamma_z V^\dagger$ in a block-diagonal form by rearranging the order of the states, where one block contains all the perturbed bulk the states and the other one contains the remainder (resulting in a block of all zeros). One can see from this rearrangement that only the states that are perturbed need to be included in the $lm$ summation. Thus, Eq.~\eqref{eqn:deltaG_term} can equivalently be written as 
\begin{align}
    \delta \mathcal{G}^{jk}_z &= \sum_{lm \in \textrm{pert}} G^{jl}_{z} \left[ \left(\tilde{\Delta} + \tilde{V}\Gamma_{z} \tilde{V}^\dagger\right) \right.
    \nonumber
    \\
    &\times \left. \left[1-\tilde{G}_{z}\left(\tilde{\Delta}+ \tilde{V}\Gamma_{z} \tilde{V}^\dagger\right)\right]^{-1} \right]_{lm} G^{mk}_{z} \,,
    \label{eqn:deltaG_simple}
\end{align}
where the tilde indicates that only the elements corresponding to the perturbed states are retained, substantially reducing the computational complexity. Note that a particular state is included in both $\tilde{V}$ and $\tilde{\Delta}$ even if it is perturbed by only one of the terms. Following a similar line of reasoning for the impurities, we get
\begin{equation}
   \Gamma_z^\mathrm{Full}= \Gamma_{z}+ \Gamma_{z}\tilde{V}^\dagger \tilde{G}_{z}\left[1 -\left( \tilde{\Delta} + \tilde{V}\Gamma_{z} \tilde{V}^\dagger\right)\tilde{G}_{z}\right]^{-1}  \tilde{V} \Gamma_{z}\,.
   \label{eqn:Gamma_full}
\end{equation}

The diagonal terms of $\mathbf{G}_z$ can be used to calculate the expected particle number from
\begin{equation}
    \rho_j = \frac{1}{\beta}\sum_{\omega_n}\left[\mathbf{G}_{i\omega_n + \mu}\right]_{jj}
    \label{eqn:rho}
\end{equation}
and to obtain the corresponding spectral function
\begin{equation}
    \mathcal{A}_j(\omega) = -2\mathrm{Im}\left[\mathbf{G}_{\omega+i0}\right]_{jj}\,.
    \label{eqn:A}
\end{equation}

From Eq.~\eqref{eqn:Z}, we can also write down the Helmholtz free energy $F = -\beta^{-1} \ln \mathcal{Z}$:

\begin{align}
    F &= -\beta^{-1}\sum_n \ln\left|\beta
    \begin{pmatrix}
        -G_{i\omega_n+\mu}^{-1} + \Delta & V
        \\
        V^\dagger & -\Gamma_{i\omega_n+\mu}^{-1}
    \end{pmatrix}
    \right|
    \nonumber
    \\
   & = -\beta^{-1}\sum_n \ln\Bigg|\beta
    \begin{pmatrix}
        -G_{i\omega_n+\mu}^{-1} & 0
        \\
        0 & -\Gamma_{i\omega_n+\mu}^{-1}
    \end{pmatrix}
    \nonumber
    \\
    &\times
    \left[
    1
    +
    \begin{pmatrix}
        -G_{i\omega_n+\mu} & 0
        \\
        0 & -\Gamma_{i\omega_n+\mu}
    \end{pmatrix}
    \begin{pmatrix}
        \Delta & V
        \\
        V^\dagger & 0
    \end{pmatrix}
    \right]
    \Bigg|\,.
    \label{eqn:F}
\end{align}

Removing the part of $F$ corresponding to the free energy of the two isolated systems in the absence of any perturbation yields the defect- and coupling-induced modification to $F$

\begin{align}
    \delta F 
   & = -\beta^{-1}\sum_n \ln\left|
    \begin{pmatrix}
       1- G_{i\omega_n+\mu}\Delta & -G_{i\omega_n+\mu}V
        \\
       - \Gamma_{i\omega_n+\mu}V^\dagger & 1
    \end{pmatrix}
    \right|
    \nonumber
    \\
    &= -\beta^{-1}\sum_n \ln\left|
      1- \tilde{G}_{i\omega_n+\mu}\left(\tilde{\Delta}+\tilde{V}\Gamma_{i\omega_n+\mu}\tilde{V}^\dagger\right)
    \right|\,.
    \label{eqn:delta_F}
\end{align}

Next, we will show how Eqs.~\eqref{eqn:rho}-\eqref{eqn:delta_F} are used to efficiently calculate interaction energy and electronic density in graphene in the presence of defects and impurities.

\subsection{Graphene Green's Function}
\label{sec:Graphene_Greens_Function}

With the formalism established, we now simply need to obtain the relevant Green's functions for the system of interest and plug them into the expressions above. Because the electronic properties of graphene are dominated by the carbon $\pi$ orbitals, the electronic states in a pristine system can be described by $|\mathbf{r},L\rangle\otimes|\sigma\rangle$, where $\mathbf{r}$ is the coordinate of the unit cell hosting the orbital, $L$ is the sublattice of the atom, and $\sigma$ is the spin of the electron. Naturally, the infinitely-large monolayer corresponds to the bulk component in the discussion above so that the $|\mathbf{r},L\rangle\otimes|\sigma\rangle$ basis refers to the operators $c_k$ in Eq.~\eqref{eqn:H}.

To calculate the matrix elements of the graphene Green's function $G_z = (z - H)^{-1}$ in the position basis, given by  $\langle\mathbf{r}, L|\otimes\langle\sigma|(z - \hat{H})^{-1}|\mathbf{r}',L'\rangle\otimes|\sigma'\rangle$, it is useful to Fourier-transform the real-space states to get
\begin{equation}
    \frac{1}{N}\sum_{\mathbf{qq}'}\langle\mathbf{q}, L| \otimes \langle \sigma|e^{i\mathbf{q}\cdot\mathbf{r}}(z - \hat{H})^{-1}e^{-i\mathbf{q}'\cdot\mathbf{r}'}|\mathbf{q}',L'\rangle\otimes|\sigma'\rangle \, ,
\end{equation}
where the $\mathbf{q}$ and $\mathbf{q}'$ momentum sums run over the entire Brillouin zone and $N$ is the number of states in the system. Because the momentum-space Hamiltonian is diagonal in $\mathbf{q}$ and $\sigma$, we can write the matrix elements as 
\begin{align}
    & \frac{1}{N}\sum_{\mathbf{q}}\langle\mathbf{q}, L|e^{i\mathbf{q}\cdot\mathbf{r}}(z - \hat{H})^{-1}e^{-i\mathbf{q}\cdot\mathbf{r}'}|\mathbf{q},L'\rangle \delta_{\sigma\sigma'} \nonumber \\
    =& \langle L|\frac{1}{N}\sum_{\mathbf{q}}e^{i\mathbf{q}\cdot\left(\mathbf{r}-\mathbf{r}'\right)}(z - H_\mathbf{q})^{-1}|L'\rangle \delta_{\sigma\sigma'} \label{final_matrix_elem}\, .
\end{align}

The graphene tight-binding Hamiltonian with the nearest-neighbor hopping is given by 
\begin{equation}
    H_\mathbf{q} =\begin{pmatrix} 0 & -t f_\mathbf{q} \\ -t f^*_\mathbf{q} & 0\end{pmatrix} \, ,
\end{equation}
where $t = 2.8 \,\textrm{eV}$ is the hopping integral, $f_\mathbf{q} = 1 + e^{i\mathbf{q}\cdot\mathbf{d}_1} + e^{i\mathbf{q}\cdot\mathbf{d}_2}$, and $\mathbf{d}_{1/2}=d\left(\pm1, \sqrt{3}\right)/2$ are the lattice vectors. From Eq.~\eqref{final_matrix_elem}, we have 
\begin{align}
    & \frac{1}{N}\sum_\mathbf{q} \left(z -  H_\mathbf{q} \right)^{-1} e^{i\left(\mathbf{r}_k - \mathbf{r}_j\right)\cdot\mathbf{q}} \nonumber \\
    =& \frac{1}{N}\sum_\mathbf{q}
    \begin{pmatrix}
    z& - tf_\mathbf{q}
    \\
    - tf_\mathbf{q}^* & z
    \end{pmatrix}\frac{1}{z^2 - t^2 \left|f_\mathbf{q}\right|^2} e^{i\mathbf{r}_{kj}\cdot\mathbf{q}} \, ,
\end{align}
where $\mathbf{r}_{jk} = \mathbf{r}_j - \mathbf{r}_k$. To perform the summation over $\mathbf{q}$, we first introduce
\begin{equation}
    \Omega^{u,v}_z =
	\frac{1}{N}\sum_{\mathbf{q}\in\mathrm{BZ}}
	\frac{
		e^{i\mathbf{q}\cdot \left(u\mathbf{d}_1 + v\mathbf{d}_2\right)}
	}
	{z^2 - t^2\left| f_{\mathbf{q}}\right|^2}
\end{equation}
with $u\mathbf{d}_1 + v\mathbf{d}_2 = \frac{d}{2}\left(u - v, \sqrt{3}\left(u+v\right)\right)$. Writing $\mathbf{q}\cdot \left(u\mathbf{d}_1 + v\mathbf{d}_2\right)  = \frac{d}{2}\left[\left(u - v\right)q_x + \sqrt{3}\left(u+v\right)q_y\right]$ and turning the momentum sum into an integral yields
\begin{equation}
    \Omega^{u,v}_{z}
	= \oint \frac{dx}{2\pi} \oint \frac{dy}{2\pi}
	\frac
	{e^{i \left[\left(u - v\right)x + \left(u+v\right)y\right]}}
	{z^2 - t^2\left(1 + 4\cos^2 x + 4 \cos x\cos y \right)}\,.
	\label{eqn:Omega_int}
\end{equation}

Using 
\begin{equation}
    \oint d\theta \frac{e^{il\theta}}{W-\cos\theta} = 2\pi \frac{\left(W - \sqrt{W - 1}\sqrt{W + 1}\right)^{|l|}}{\sqrt{W - 1}\sqrt{W + 1}}\,,
\end{equation}
turns Eq.~\eqref{eqn:Omega_int} into
\begin{equation}
    \Omega^{u,v}_z = 
	\oint \frac{dx}{2\pi} \frac{e^{i\left(u - v\right)x}}{\cos x} \frac{\left(W - \sqrt{W - 1}\sqrt{W + 1}\right)^{|u+v|}}{4t^2\sqrt{W - 1}\sqrt{W + 1}}\,,
\end{equation}
where $W = (z^2 - t^2)/(4 t^2 \cos x) - \cos x$. Finally, for $\mathbf{r} = u\mathbf{d}_1 + v\mathbf{d}_2$, we have
\begin{align}
    &\frac{1}{N}\sum_\mathbf{q} \left(z -  H_\mathbf{q} \right)^{-1} e^{i\mathbf{r}\cdot\mathbf{q}}  \nonumber \\
	&=
	\begin{pmatrix}
		z\Omega^{u,v}_z
		&
		- t\left[\Omega^{u,v}_z + \Omega^{u,v}_{+,z} \right]
		\\
		- t\left[\Omega^{u,v}_z + \Omega^{u,v}_{-,z}\right]
		&
		z\Omega^{u,v}_z
	\end{pmatrix}\,,
\end{align}
where 
\begin{equation}
    \Omega^{u,v}_{\pm,z}
	=
	\oint \frac{dx}{2\pi} \,2e^{i\left(u - v\right)x} \frac{\left(W - \sqrt{W - 1}\sqrt{W + 1}\right)^{|u+v\pm 1|}}{4t^2\sqrt{W - 1}\sqrt{W + 1}}\,.
\end{equation}
The one-dimensional integrals over $x$ can be computed efficiently using Gaussian quadratures.

Using the multiples of the basis vectors to describe the electronic states, the matrix elements $\langle u, v, L|\otimes\langle\sigma|(z - \hat{H})^{-1}|u', v' ,L'\rangle\otimes|\sigma'\rangle$ of the Green's function become
\begin{widetext}
\begin{equation}
   \langle u, v, L|\otimes\langle\sigma|(z - \hat{H})^{-1}|u', v' ,L'\rangle\otimes|\sigma'\rangle= \langle L|\begin{pmatrix}
		z\Omega^{u-u',v-v'}_z
		&
		- t\left[\Omega^{u-u',v-v'}_z + \Omega^{u-u',v-v'}_{+,z} \right]
		\\
		- t\left[\Omega^{u-u',v-v'}_z + \Omega^{u-u',v-v'}_{-,z}\right]
		&
		z\Omega^{u-u',v-v'}_z
	\end{pmatrix}|L'\rangle \delta_{\sigma\sigma'}
\end{equation}
\end{widetext}
with $|L'\rangle$ and $\langle L|$ picking out the appropriate element of the matrix depending on the sublattices of the two states.

\subsection{Hopping and Spin Defects}
\label{sec:Defects}

The $\Delta$ matrix introduced in Sec.~\ref{sec:Two_Component_System} describes the modified coupling between the states in the bulk. In the case of graphene, it can be used to encode an on-site potential, a modified hopping term, and an interaction between graphene's electrons and localized spin moments. In the first case, $\Delta$ acquires diagonal terms for both spins, while for the hopping modification, $\Delta$ gets identical off-diagonal terms for the two spins.

To encode the coupling between graphene electrons and localized spins, recall that the spin-spin interaction can be written as $\mathbf{s}_{i}\cdot J_{ik}\cdot\hat{\boldsymbol{\sigma}}_{\sigma\sigma'}c_{k\sigma}^\dagger c_{k\sigma'}$, where $\mathbf{s}_i$ is the $i^{th}$ localized spin moment and $J_{ik}$ is the coupling constant between the $i^{th}$ spin and $k^{th}$ electronic state. We write $\mathbf{S}_{ik}=\mathbf{s}_i \times J_{ik}$ as the effective coupling strength between the spin and the carbon atom into which we absorbed the interaction strength and the spin angular momentum so that it has the units of energy. Writing out this expression for graphene yields
\begin{align}
    \mathbf{S}_{jk}\cdot\hat{\boldsymbol{\sigma}}_{\sigma\sigma'}c_{\sigma,\mathbf{R}_j}^\dagger c_{\sigma',\mathbf{R}_j}& = (S_z)_{jk}\left(c_{\uparrow,\mathbf{R}_j}^\dagger c_{\uparrow,\mathbf{R}_j}-c_{\downarrow,\mathbf{R}_j}^\dagger c_{\downarrow,\mathbf{R}_j}\right) 
    \nonumber
    \\
   & +
    \Big((S_x)_{jk} - i (S_y)_{jk}\Big)c_{\uparrow,\mathbf{R}_j}^\dagger c_{\downarrow,\mathbf{R}_j}
    \nonumber
    \\
    &+
    \Big((S_x)_{jk} + i (S_y)_{jk}\Big)c_{\downarrow,\mathbf{R}_j}^\dagger c_{\uparrow,\mathbf{R}_j}\,,
    \label{eqn:Spin_Delta}
\end{align}
where all the $c$ operators belong to either A or B sublattice. One can see that $S_z$ plays the role of a spin-dependent on-site potential, entering $\Delta$ as a diagonal term and the planar components of the localized spin give the hopping between the two spin orbitals of a carbon atom.

Treating the hopping perturbation on the same footing as the interaction with the localized spins makes it possible to simultaneously treat different defect types following the procedure outlined in Sec.~\ref{sec:Two_Component_System}.

\section{Results}
\label{sec:Results}
With the details of the formalism outlined in Sec.~\ref{sec:Model}, we can now use GrapheneQFT.jl to explore various defect configurations in graphene. Assuming that the user has Julia installed on their system, the package is installed and imported in the usual manner:
\begin{minted}[bgcolor = light-gray]{julia}
using Pkg; Pkg.add("GrapheneQFT")
using GrapheneQFT
\end{minted}

The key object from which all physical quantities are computed is a \mintinline{julia}{GrapheneSystem}. This object, containing the system's chemical potential, temperature, and the $\Delta$ and $V$ matrices defined in Sec.~\ref{sec:Two_Component_System}, can be initialized using \mintinline{julia}{mkGrapheneSystem(μ, T, defects)}, where \mintinline{julia}{defects} is an array of defects in the system. The defects can be of one of three types: localized states (corresponding to the impurity subsystem in Sec.~\ref{sec:Model}), a localized spin state, or a hopping modification. Below, we demonstrate how one can study the effects of localized spins and refer the reader to the package documentation~\citep{rodinalex_harshitra-m_2022} for more examples dealing with other defect types.

\subsection{Local Electronic Density}
\begin{figure*}[ht]
    \centering
    \includegraphics[width=\textwidth]{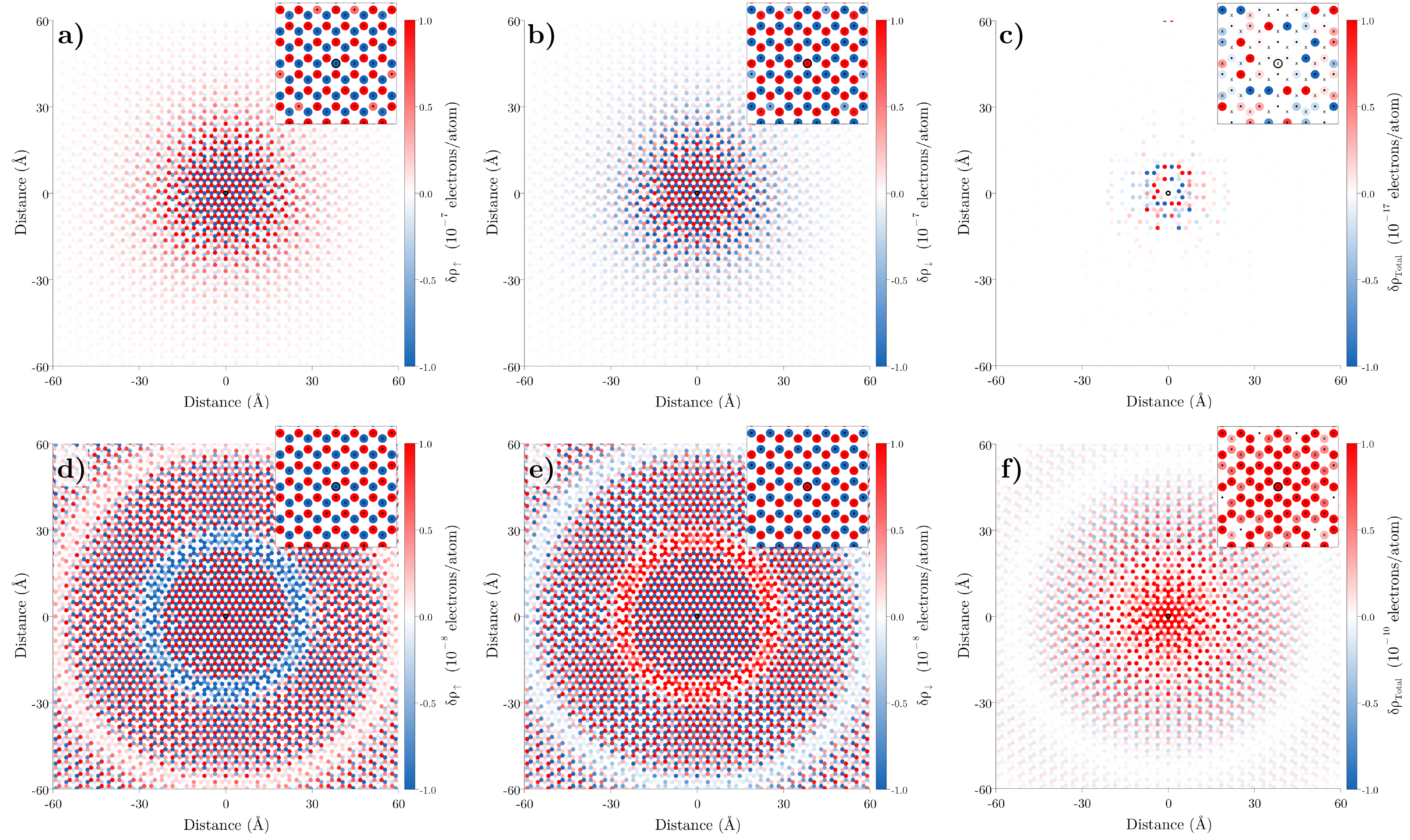}
    \caption{\textbf{Spin-resolved and total electron density variation.} From left to right: variation in spin-up, spin-down and total electron density for a system with a single magnetic impurity. The top row shows the variation at $\mu = 0.0$ eV, while the bottom row shows the variation at $\mu = 0.3$ eV. The impurity is located at the center of system (marked with black outline) with spin triplet $\mathbf{S} = J(0,0,1)$, where the spin-lattice coupling is set to $J = 0.01$ eV. The inset plots show the electron density variation around the area close to impurity. Note that the color scale is saturated to demonstrate the sublattice dependence and the scale between the spin-resolved and total electron density varies greatly.}    
    \label{fig:EDV}
\end{figure*}
Consider a system with a single localized spin defect pointing in the $z$-direction, coupled to the A-sublattice atom at the unit cell with $u = v = 0$. We first create a \mintinline{julia}{GrapheneSystem} containing this spin:

\begin{minted}[bgcolor = light-gray, breaklines]{julia}
# System parameters
μ = 0.0
T = 0.0

# Spin-lattice coupling
J_val = 0.01

# Single spin defect
single_spin_sys = mkGrapheneSystem(μ, T, Defect[LocalSpin(0.0, 0.0, J_val, GrapheneCoord(0, 0, A))])
\end{minted}
The single entry in the \mintinline{julia}{Defect[]} array is the desired \mintinline{julia}{LocalSpin} with $S_x = S_y = 0$ and $S_z = J_\mathrm{val}$, coupled the graphene coordinate $(u = 0, v = 0, \mathrm{A})$. The function \mintinline{julia}{mkGrapheneSystem} performs all the necessary manipulations to generate the required $\Delta$ and $V$ matrices.

Next, we calculate the defect-induced density. The local electronic density variation induced by the presence of defects, $\delta \rho_\mathbf{R}$, can be calculated similarly to Eq.~\eqref{eqn:rho}, except with the diagonal terms of $\mathbf{\delta G}_{i\omega_n + \mu}$. The package provides a function, making this calculation straightforward. For example, \begin{minted}[bgcolor = light-gray, breaklines]{julia}
δρ_R_graphene(GrapheneState(GrapheneCoord(4,5,B), SpinUp), single_spin_sys)
\end{minted}
calculates the defect-induced density at $|4,5,\mathrm{B}\rangle\otimes|\uparrow\rangle$ for the system defined above. Using the BenchmarkTools.jl library~\citep{BenchmarkTools.jl-2016} to benchmark the function for this particular system, the mean runtime for a single call (averaged over 1413 samples) is $3.532$ms $\pm \, 883.068 \mu$s.
\begin{figure*}[ht]
    \centering
    \includegraphics[width = \textwidth]{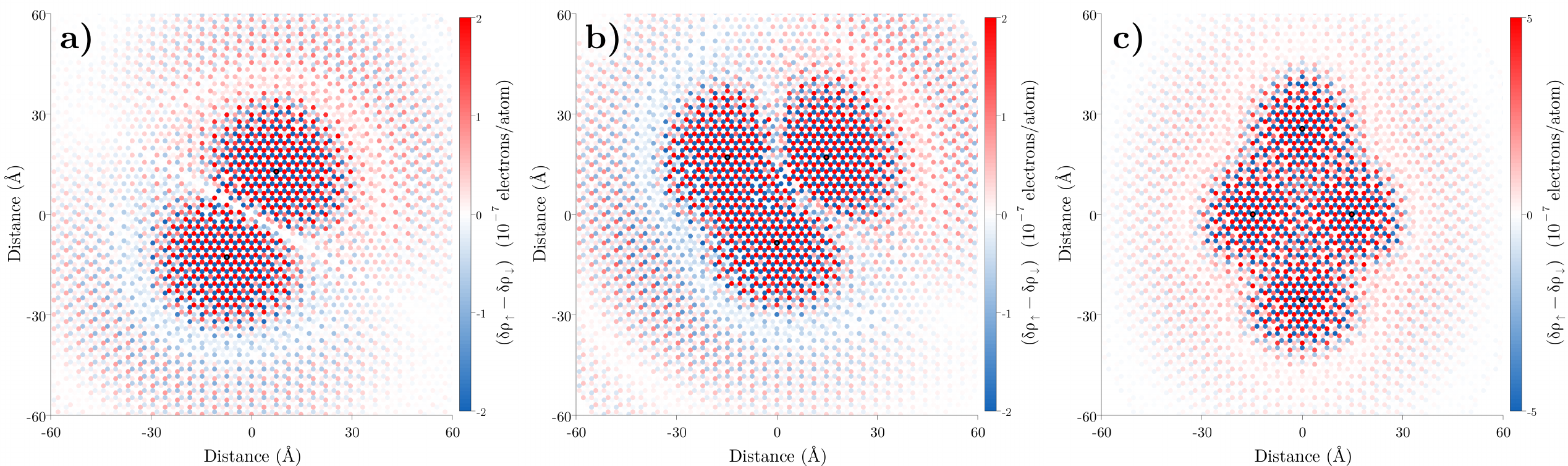}
    \caption{\textbf{Spin asymmetry $\delta \rho_\uparrow - \delta \rho_\downarrow$ in multi-impurity systems.} From (a) to (c), the variation in spin asymmetry $(\delta \rho_\uparrow - \delta \rho_\downarrow)$ for systems with two, three and four localized spins, respectively. The localized spins (marked with black outline) are positioned with a distance of $12d$ between each of them, where $d$ is the graphene lattice constant and have a spin-lattice coupling of $J = 0.01$ eV. All plots are at $\mu = 0.3$ eV. (a) shows two spins in the antiferromagnetic configuration, (b) shows three spins in a frustrated triplet configuration (up-up-down) and (c) shows four spin-downs.}
    \label{multi_imp}
\end{figure*}
To produce a density map, we define a list of coordinates for which $\delta \rho_\mathbf{R}$ is calculated
\begin{minted}[bgcolor = light-gray, breaklines]{julia}
# Grid parameters
nPts = 50
d1s = -nPts:1:nPts
d2s = -nPts:1:nPts

# Coordinate grids
coord_A = [GrapheneCoord(u,v,A) for u in d1s, v in d2s] |> vec
coord_B = [GrapheneCoord(u,v,B) for u in d1s, v in d2s] |> vec
\end{minted}
Applying \mintinline{julia}{δρ_R_graphene} to every coordinate in \mintinline{julia}{coord_A} and \mintinline{julia}{coord_B} for both spins produces arrays of corresponding densities. GrapheneQFT.jl also provides a function \mintinline{julia}{crystal_to_cartesian} which converts crystal coordinates in \mintinline{julia}{coord_A} and \mintinline{julia}{coord_B} to Cartesian ones, making it straightforward to plot the induced charge densities as scatter plots, as shown in Fig.~\ref{fig:EDV}.

The localized spin oriented in the $z$-direction effectively creates a spin-dependent local potential for the electrons in graphene. When the chemical potential $\mu = 0.0$ eV, as it is in the top row of Fig.~\ref{fig:EDV}, the induced density for spin-up [panel (a)] and spin-down [panel (b)] is equal in magnitude and opposite in sign, as can be seen from panel (c), where the two densities are summed. The finite values in Fig.~\ref{fig:EDV}(c) are the consequence of the finite numerical precision, as one can observe from their magnitude.

Raising $\mu$ to $0.3$ eV demonstrates another effect that the symmetry-breaking scatterer has on the system: the Friedel oscillations, seen in the bottom row of Fig.~\ref{fig:EDV}. Earlier work~\citep{Lawlor2013, Noori2020} discussed the sublattice dependence of the Friedel oscillations in graphene, noting that the two sublattices have the same oscillation period, but their phase is shifted with respect to each other. One can see in Fig.~\ref{fig:EDV}(d)-(e) that the two sublattices typically have opposite signs of $\delta\rho_\mathbf{R}$. When the sign of the two sublattices coincides, we observe regions of charge accumulation and depletion, seen as the red and blue circles in Fig.~\ref{fig:EDV}(d)-(e). As expected, the signs are reversed between the spin-up and spin-down electrons.

For $\mu=0.3$ eV, the Fermi momentum $k_F \approx 0.046$ \AA$^{-1}$. The corresponding wavelength of the Friedel oscillation is $\pi/k_F \approx 68$ \AA. The difference between the radii of the blue and red circles ($\sim 33$ \AA) correspond to the half-wavelength of the Friedel oscillations, in agreement with the expected value.

One of the unique features of our code is the ability to analyze any number of impurities in any configuration, as seen in Fig.~\ref{multi_imp}. Fig.~\ref{multi_imp}(a)-(c) show the spin asymmetry variation $(\delta\rho_\uparrow - \delta\rho_\downarrow)$ for systems with two, three and four equidistant localized spins oriented in the $z$-direction. In panel (a) the two impurities have opposite spin directions, leading to density variations with opposite signs immediately around the impurities. These variations cancel out at regions equidistant from both impurities, leading to a suppression of the electron density variation in the region between the localized spins. In (c) all 4 spins are ferromagnetically aligned with each other and cause constructive interferences between variations from individual impurities. Panel (b) has 3 localized spins in a frustrated configuration, with two spin-up impurities and a spin-down impurity. As the spin-down impurity is located on the top right corner of the triangle, there is suppression of the electronic density in the region equidistant from the two spin-ups and the spin-down.

\subsection{Interaction Energy}

\begin{figure*}[ht]
    \centering
    \includegraphics[width = \textwidth]{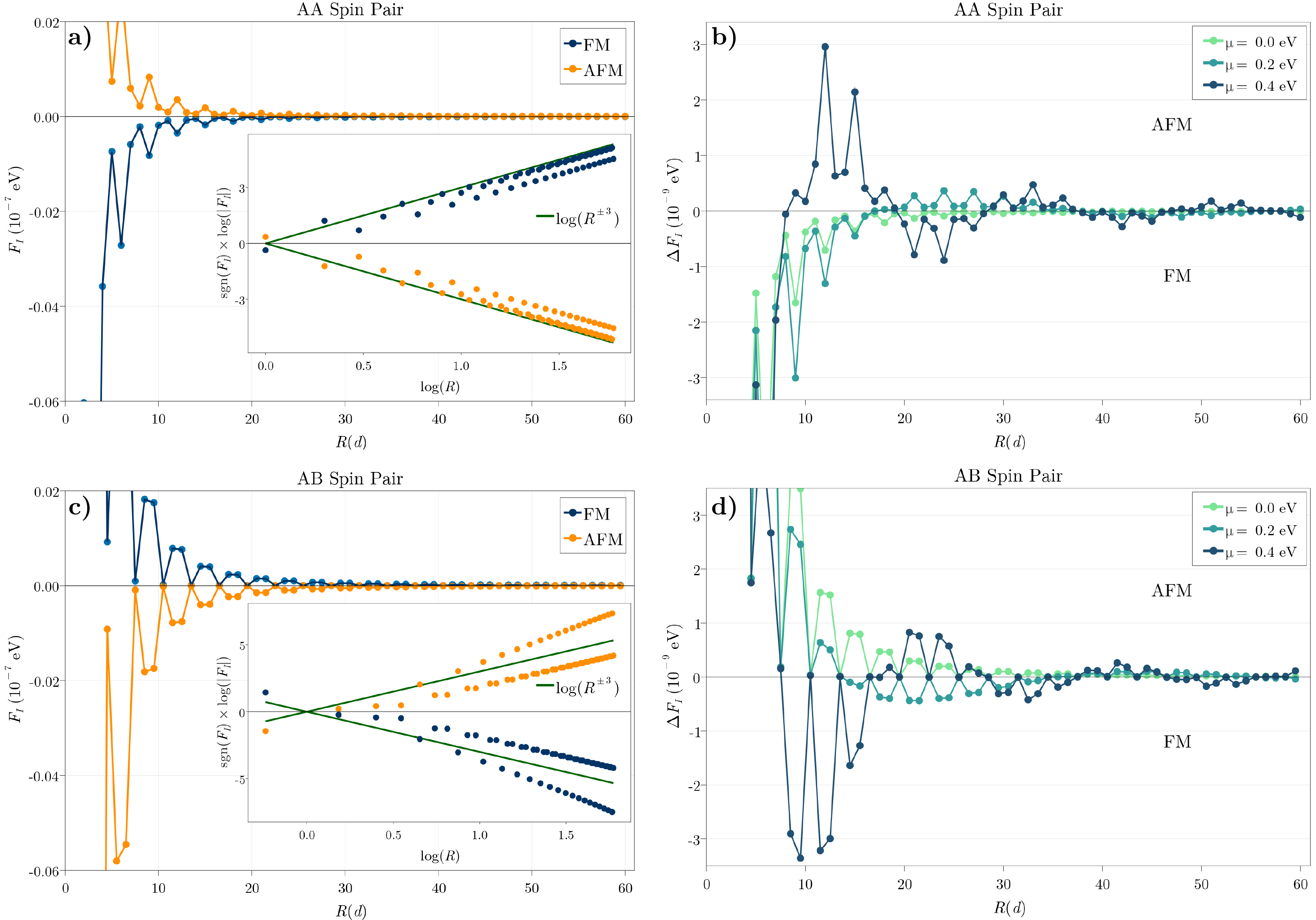}
\caption{\textbf{Interaction energy between localized spin pairs separated along the zigzag direction.} From top to bottom, the interaction energy $F_I$ and $\Delta F_I = F_I^{\mathrm{FM}} - F_I^{\mathrm{AFM}}$ for spin pairs positioned on the same sublattice (AA) and on opposite sublattices (AB). (a) and (c) show $F_I$ for undoped graphene with the inset plots showing the log-log plot. Depending on the configuration, the spin pairs have spin triplets $\mathbf{S} = \pm J(0.5, 0.0, -\sqrt{3}/2)$, where the spin-lattice coupling is $J = 0.01$ eV. The difference between $F_I$ for FM and AFM configurations is plotted in b) and d) for different doping levels.}
\label{spin_pair_zigzag}
\end{figure*}
\begin{figure*}[ht]
    \centering
    \includegraphics[width = \textwidth]{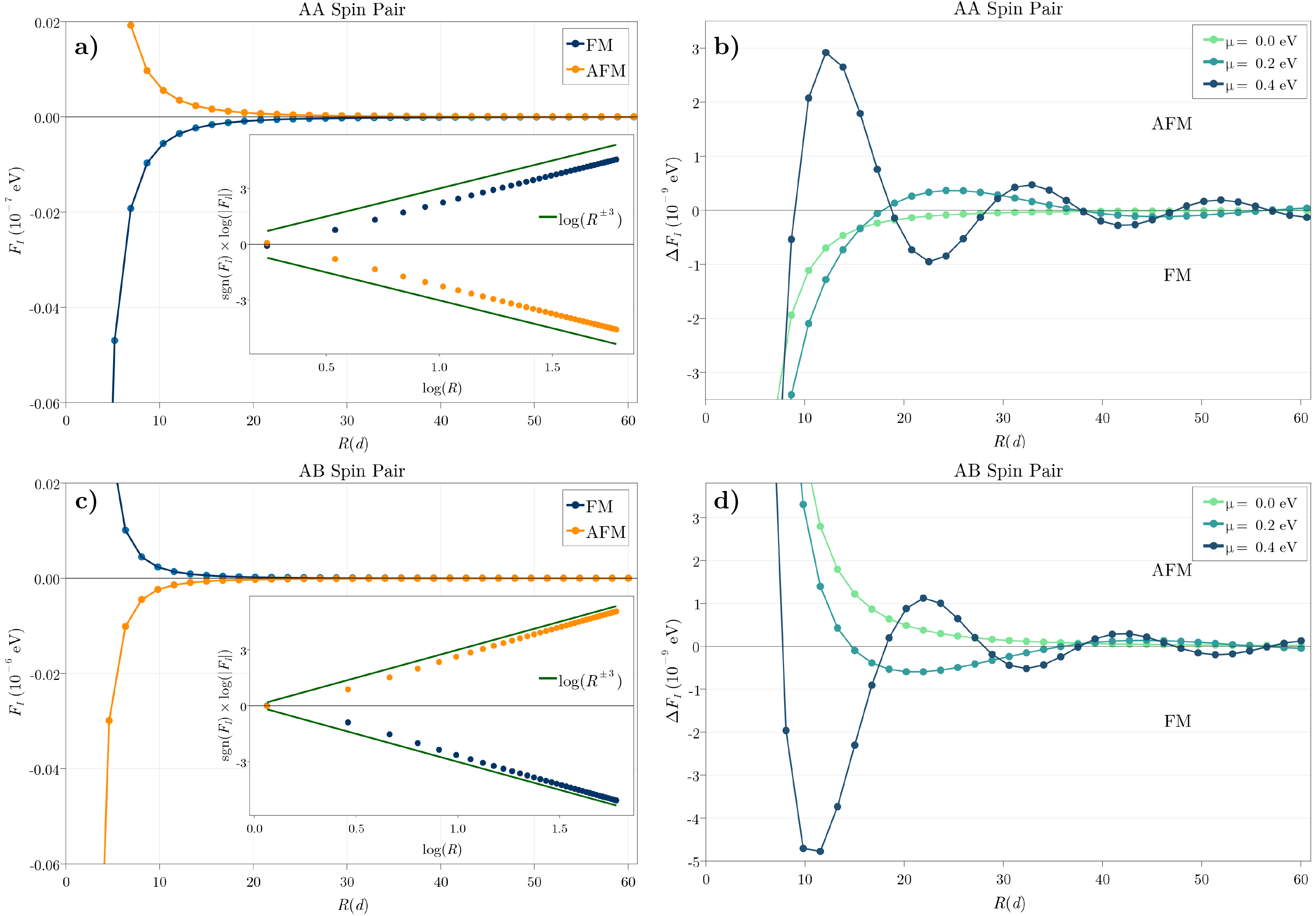}
    \caption{\textbf{Interaction energy between localized spin pairs separated along the armchair direction.} From top to bottom, the interaction energy $F_I$ and $\Delta F_I = F_I^{\mathrm{FM}} - F_I^{\mathrm{AFM}}$ for spin pairs positioned on the same sublattice (AA) and on opposite sublattices (AB). (a) and (c) show $F_I$ for undoped graphene with the inset plots showing the log-log plot. Depending on the configuration, the spin pairs have spin triplets $\mathbf{S} = \pm J(0.5, 0.0, -\sqrt{3}/2)$, where the spin-lattice coupling is $J = 0.01$ eV. The difference between $F_I$ for FM and AFM configurations is plotted in b) and d) for different doping levels.}
    \label{spin_pair_armchair}
\end{figure*}

As discussed above, $\delta F$ in Eq.~\eqref{eqn:delta_F} gives the variation in free energy due to the introduction of impurities.  From there, interaction energy among impurities can be calculated in the following way:
\begin{equation}
    F_{I} = \delta F^{(n)} - \sum_j \delta F_j^{(1)} \, ,
    \label{eqn:F_I}
\end{equation}
where $\delta F^{(n)}$ refers to the defect-induced variation in energy in the $n$-impurity system in consideration and $\delta F_j^{(1)}$ refers to the defect-induced energy variation for a single impurity $j$. To calculate $F_I$ between two impurities, we build upon the \mintinline{julia}{δF} function in GrapheneQFT.jl and define a function tailored to the system we are interested in. For example, the function below uses Eq.~\eqref{eqn:F_I} to calculate the interaction energy between two anti-parallel localized spins separated along the zigzag direction:
\begin{minted}[bgcolor = light-gray, breaklines, mathescape, escapeinside = ||]{julia}
function Fint_spin_pair_AA_zz(sep::Int64)
    # Localized spins
    coords = [GrapheneCoord(0,0,A), GrapheneCoord(sep,0,A)]
    spin1 = LocalSpin(0.0, 0.0, J_val, coords[1])
    spin2 = LocalSpin(0.0, 0.0, -J_val, coords[2])
    
    # Energy variation by two spins
    two_spins = δF(mkGrapheneSystem(μ, T, Defect[spin1, spin2])) 
    
    # Energy variation by single spin
    single_1 = δF(mkGrapheneSystem(μ, T, Defect[spin1]))
    single_2 = δF(mkGrapheneSystem(μ, T, Defect[spin2]))
    
    return (two_spins-(single_1+single_2))
end
\end{minted}
Using BenchmarkTools.jl, we see that the averaged runtime for a single call at separation distance $8d$ (averaged over 741 samples) is $6.743$ ms $\pm \,  1.281$ ms.

In Fig.~\ref{spin_pair_zigzag} and \ref{spin_pair_armchair}, we calculate and plot the variation in interaction energy between two impurities separated by distance $R$ along the zigzag and armchair directions, respectively. For undoped graphene [panels (a) and (c) in Fig.~\ref{spin_pair_zigzag} and \ref{spin_pair_armchair}], ferromagnetic (anti-ferromagnetic) configuration is preferred for the same (opposite) sublattice configuration. Depending on the configuration, the inner envelope of the interaction energy follows a $\pm R^{-3}$ power law decay [inset plots of Fig.~\ref{spin_pair_zigzag} and~\ref{spin_pair_armchair}], in line with Refs.~\citep{Saremi2007rih, Agarwal2017lre}.

We also plot $\Delta F_I = F_I^\mathrm{FM} - F_I^\mathrm{AFM}$ for several values of $\mu$ in panels (b) and (d) of Fig.~\ref{spin_pair_zigzag} and~\ref{spin_pair_armchair}, where $\Delta F_I<0$ indicates ferromagnetic ordering. It can be considered as an indicator of spin-spin interaction strength. We can see for undoped systems, same-sublattice (AA) configurations give rise to ferromagnetic ordering, while different-sublattice (AB) arrangements produce anti-ferromagnetic spin orientations regardless of the direction along which the spins are positioned.

In Fig.~\ref{spin_pair_zigzag} we see an additional short-range oscillation of wavelength $3d = 7.38$\AA\ that makes up the outer envelope of the interaction energy decay profile and is independent of the chemical potential of the system. This is a signature of a system with multivalley band structure \cite{PhysRevB.45.1660} and it arises from intervalley scattering between two valleys of graphene (K and K'). $|\mathbf{K}-\mathbf{K}'| = \frac{4\pi}{3d}$, hence the actual wavelength is $2\pi/|\mathbf{K}-\mathbf{K}'|=3d/2$. It is manifested in the coupling strength variation at every $3d$ increment in impurity separation. The inclusion of the whole Brillouin zone allows us to capture this subtle variation in coupling strength. These oscillations are not present when spin pairs are separated along the armchair direction (Fig.~\ref{spin_pair_armchair}). Akin to the aliasing effect in signal sampling, the spin separation of $\sim 1.73 d$ is larger than the wavelength of the short-range oscillation and thus does not affect the coupling strength variation.

As we increase the chemical potential, we see that $\Delta F_I$ is not always negative (positive) for the AA (AB) configuration anymore. The Friedel oscillations that occur with non-zero chemical potential results in $\Delta F_I$ crossing zero multiple times, depending on the value of $\mu$. This gives rise to regions with ferromagnetic and antiferromagnetic ordering that is dependent on the separation distance between the two localized spins. Since the Friedel wavelengths decrease as $\mu$ increases, the sublattice dependence reversal happens more and more frequently at higher $\mu$ values ($\mu=0.2$ eV and $\mu=0.4$ eV curves in the (b) and (d) panels of Fig.~\ref{spin_pair_zigzag} and~\ref{spin_pair_armchair}).

\begin{figure}[ht]
    \centering
    \includegraphics[width = 0.48\textwidth]{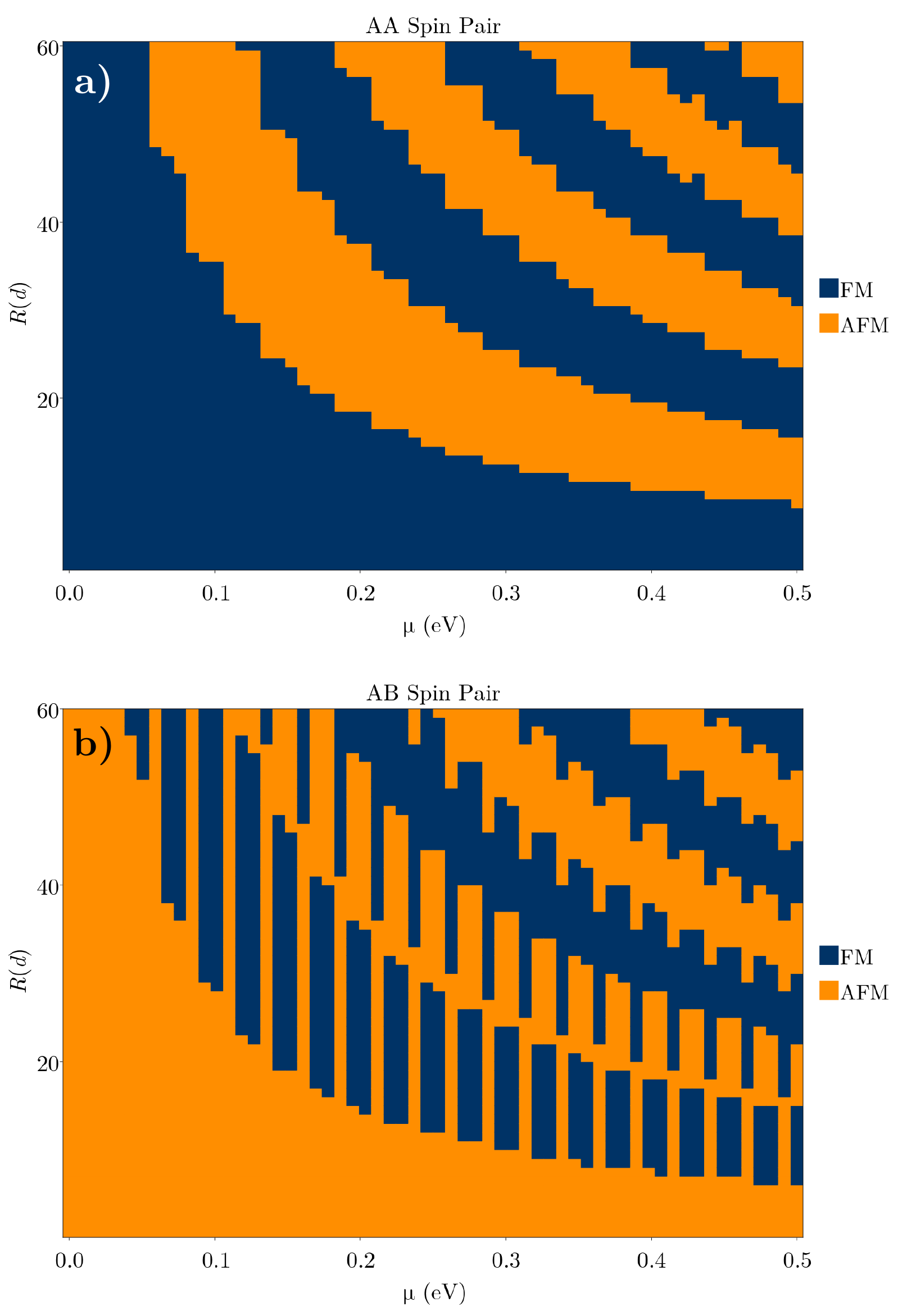}
    \caption{\textbf{Preferred ordering of localized spin pairs as a function of separation distance and $\mu$.} Variation in preferred ordering of spin pairs hosted by (a) atoms of the same sublattice and (b) atoms of different sublattices as the separation distance between the two localized spins and the doping levels are varied.}
    \label{spin_pair_heatmap}
\end{figure}

\subsection{Magnetic Order}
The quantity $\Delta F_I$ allows us to predict the magnetic order of a particular impurity arrangement. In agreement with the available literature, our calculations show that the spin orientation depends on the sublattice arrangement, the distance between the impurities, and the chemical potential of the system. At small separations, same-sublattice configurations favor ferromagnetic orientation, while the opposite-sublattice arrangements favor anti-ferromagnetic orientation. At larger separations, $\Delta F_I$ exhibits an oscillatory behavior, leading to an alternation of ferromagnetic/anti-ferromagnetic order for both sublattice configurations, as shown in panels (b) and (d) of Fig.~\ref{spin_pair_zigzag} and ~\ref{spin_pair_armchair}. The wavelength of the oscillations is related to the doping level as they are a manifestation of the Friedel oscillations. We demonstrate the interplay between the spin separation and the system doping in Fig.~\ref{spin_pair_heatmap}.

The fact that a particular sublattice configuration (either AA or AB) can result in both FM or AFM coupling produces an unique opportunity where we can create frustration among the impurities by placing them at certain positions. For example, let us place three impurities on the same sublattice at the vertices of an equilateral triangle. Now, if the length of the side of the triangle falls within a preferred range where all three impurities prefer to be aligned anti-ferromagnetically, the impurities will be frustrated. If we change $\mu$ gradually, at some point the impurities will favor the ferromagnetic alignment and the frustration will disappear. Hence, we can essentially turn the frustration on and off by varying the doping level of the system.
 
When considering different configurations, the computational framework lends itself well to the application of optimization algorithms. Indeed, the choice of spin triplet $\mathbf{S}$ in Fig.~\ref{spin_pair_zigzag} and~\ref{spin_pair_armchair} is motivated by allowing the spin pairs to have in-plane components and choosing the spin orientations that minimize $F_I$. In the case of the frustrated trio of impurities, the anti-ferromagnetic frustration only exists if the spins are limited to the out-of-plane orientation. Minimizing $F_I$ for three spins while allowing the spins to have in-plane components resolves the frustration and the impurity spins orient themselves in a staggered position (for our example scenario, they stay at an angle of $2\pi/3$ with each other on the same plane).

\section{Summary}
\label{sec:Summary}

We have developed an effective computation scheme for treating impurities in graphene. Our field-theoretical formulation makes is possible to study dopants, adsorbates, and spin impurities simultaneously. The syntax of our package aims to keep the learning curve as gentle as possible so that community members can use it in their research and as a learning tool.

To demonstrate the package in action, we performed a detailed analysis of spin-spin interaction between magnetic impurities. Our results capture the non trivial oscillatory behavior in impurity induced variation in charge density, impurity interaction energy and magnetic ordering. We can see that each of the aforementioned quantities has a sublattice dependence which is consistent with the existing literature. We also showed that spin frustration can be switched on and off using doping.

Given the usefulness of the field-theoretic formulation in condensed matter systems, we believe that developing a scheme not limited to a single Hamiltonian would be of great benefit to the community. Our long-term plans include coming up with appropriate abstractions that would make it possible to swap the Hamiltonian.

\section*{Acknowledgements}

This work is supported by the National Research Foundation, Prime Minister Office, Singapore, under its Medium Sized Centre Programme and the support by Yale-NUS College (through Grant No. A-0003356-42-00).

\end{document}